\documentstyle[11pt,newpasp,twoside,graphicx]{article}
\def\edcomment#1{\iffalse\marginpar{\raggedright\sl#1\/}\else\relax\fi}
\def\eck#1{\left \lbrack#1\right \rbrack}
\def\rund#1{\left(#1\right)}
\reversemarginpar

\begin{document}
\title{ALI in Rapidly Expanding Envelopes}
 \author{Peter H\"oflich}
\affil{Dept. of Astronomy, University of Texas, Austin, TX 78712, USA}
\begin{abstract}
 We discuss the current implementation of the ALI method into our
HYDrodynamical RAdiation (Hydra)
code for rapidly expanding, low density 
envelopes commonly found in core collapse and thermonuclear 
supernovae, novae and WR stars.
 Due to the low densities, non-thermal excitation by high 
energy photons (e.g. from radioactive decays) and the time 
dependence of the problem, significant departures from LTE are common 
throughout the envelope even at large optical depths. 
 
 ALI is instrumental for both the coupling of the statistical equations 
 and the hydrodynamical equations with the radiation transport (RT). 
 We employ several concepts to improve the stability, and convergence 
rate/ control including the {\sl concept of leading elements}, 
the use of net rates, level locking, reconstruction of global 
photon redistribution functions, equivalent-2-level approach, and 
predictive corrector methods. For appropriate conditions, 
the solution of the time-dependent rate equations can 
 be reduced to the time-independent problem plus the (analytic) 
solution of an ODE 
 For the 3-D problem, we solve the radiation transport via the moment equations. 
To construct the Eddington tensor elements, we use a Monte Carlo scheme to 
determine the deviation of the solution for the RT equation from the 
diffusion approximation (ALI of second kind). 
 
 At the example of a subluminous, thermonuclear supernova (SN99by), we show 
an analysis of the light curves, flux and polarization spectra and discuss 
the  limitations of our approach.
\end{abstract}
\section{Physical and Numerical    Environment}
\begin{figure}[ht]
\includegraphics[width=8.2cm,angle=270]{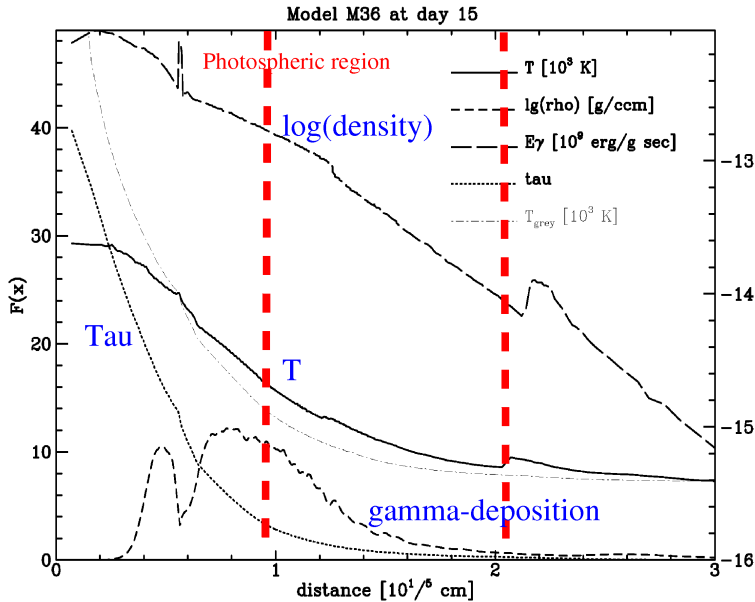}
\vskip -0.04cm
\caption{
Temperature T,  energy deposition due to radioactive
decay $E_{\gamma} $, Rosseland optical depth $Tau$ (left scale) and density log($\rho $) (right scale) are given as a function of distance
 (in $10^{15}cm$) for a typical SNe~Ia at
 15 days after the explosion. For comparison, we give the temperature
$T_{grey}$ for the grey extended atmosphere.
 The light curves and spectra of Type Ia Supernovae are powered by energy release
due to radioactive decay of $^{56}Ni \rightarrow ^{56}Co \rightarrow ^{56}Fe$. The
two dotted, vertical lines indicate the region of spectra formation. 
Most of the energy is deposited within the photosphere and, due to the small optical depth
and densities, strong NLTE effects occur up to the very central region
 (from \cite{H95}).
}
\label{sn94d}
\end{figure}

For supernovae, novae and   Wolf Rayet stars,
detailed hydrodynamical radiation calculations are required to provide a link between
the observables such as light curves and spectra, and the underlying physics of the objects. These
applications  go well beyond classical atmospheres. Density structures require detailed hydrodynamics,
the low densities cause strong NLTE effects throughout the envelopes, chemical profiles are depth
dependent,  the energy source and sink terms due to hydrodynamical effects and radioactive
decays may dominate throughout the photon decoupling region, i.e. location of the photosphere and
the physical properties are time dependent (see Fig. \ref{sn94d}). Typically, velocity fields are of the
order of 500 to 30,000 km/sec. Thus, as a major simplification, we can neglect
the intrinsic line widths due to pressure broadening, magnetic fields, etc.
\begin{figure}[ht]
\hskip 0.5cm \includegraphics[width=9.2cm,angle=270]{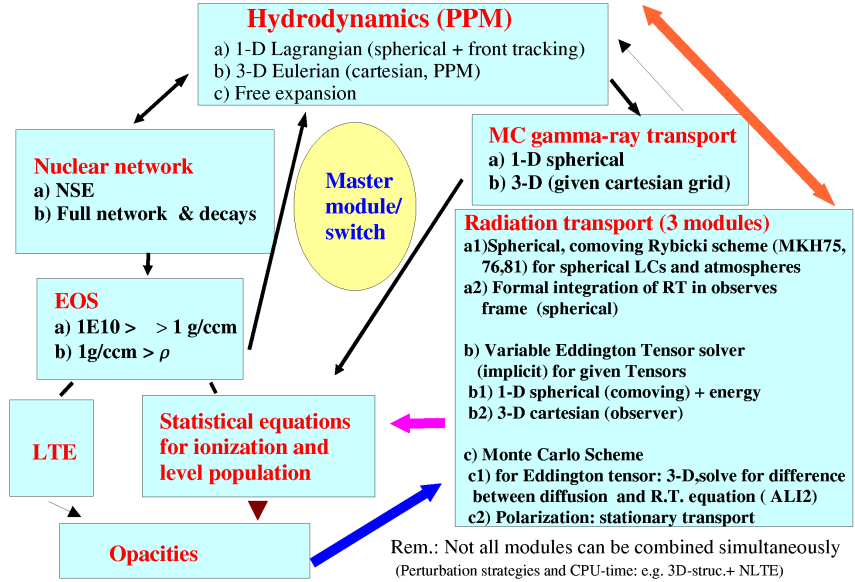}
\vskip -0.04cm
\caption{
Block diagram of our numerical scheme to solve
radiation hydrodynamical problems including detailed equation of 
state, nuclear and atomic networks. For specific problems, a subset
of the modules is employed  (see text, and Figs. \ref{SN99by} \& \ref{1dvs3d}).
The code is still under active development. For its current status, see
the applications below.
}
\label{module}
\end{figure}
 
\subsection{Numerical Tools and Methods}
The computational tools summarized below were used to carry out many of
the analyzes of SNIa and Core Collapse Supernovae ({\cite{H88}, H\"oflich, M\"uller \& Khokhlov 1993,
\cite{H95}, \cite{HHWW01}, ...}).
For the nomenclature we follow Mihalas~ \& Mihalas (1988)        if quantities are not defined explicitly.
 All components of the codes have been  written or adopted
in a  modular form with well defined interfaces which allows
an easy coupling (see Fig. \ref{module} and code verification by exchanging e.g. various
radiation transport modules but keeping identical the remaining setup (e.g. Fig. \ref{1dvs3d}).
 The modules consist of physical units to provide a solution for e.g. 
the nuclear network, the statistical equations to determine the atomic level
population, equation of states, the  opacities, 
the hydro or the  radiation transport problem.
The individual modules are coupled explicitly.
 Consistency between the solutions is achieved iteratively
by perturbation methods. Currently, not all modules can be combined simultaneously
because not all iteration schemes have been implemented and
because of requirements on CPU time:   a) For full NLTE-spectra with large model atoms
and a high frequency resolution we are restricted to the time-independent case based on
a given hydrodynamical structure (see Fig. \ref{SN99by}) and, for 3-D models, reduced atomic
atomic models  (level-merging/super-levels) have to be used  (Fig. \ref{1dvs3d}.
 b) Radiation hydrodynamics is restricted to a reduced  frequency
 resolutions with reduced atomic levels (level-merging) and spherical geometry 
(see Fig. \ref{SN99by}. In case of  multi-dimensional radiation hydrodynamics, CPU-time requirements   restrict applications even further.
Currently, we can
use of few frequencies to represent the fluxes e.g. in the Lyman and 'Balmer and higher continua', and 3-level atoms plus
spherically symmetric velocity fields for the RT({\cite{HKW02}), or to the grey case for arbitrary field.
 Technically, we use the so called 'trivial' parallelization:
 Parallelization (via MPI and PVM) is achieved on the module 
basis, e.g. parallelization of the nuclear network over the grid points,
using groups of photons for the Monte-Carlo  radiation transport,
and sub-domains with 'ghost-cells' with respect to the spacial and 
frequency coordinates  for the hydro and the comoving frame radiation transport, respectively.
 In the remainder of this section, we want to describe the various modules.

\noindent{\bf Hydrodynamics:} The structure of the expanding envelopes are obtained by three different modules
which by a) assuming free expansion, by solving the hydro equations in b)  the Lagrangian frame for
spherical geometry including a front tracking scheme to resolve shock fronts
 (e.g. \cite{HWT98}), or the Eulerian scheme for full 3-D using Cartesian coordinates based
on PROMETHEUS (\cite{FAM91}).
The hydro modules use an explicit Piecewise Parabolic Method (PPM) by \cite{CW84} to
solve the compressible reactive flow equations.  PPM is implemented as a
step followed by separate remaps of the thermal and kinetic
energy to avoid numerical generation of spurious pressure disturbances
during propagation of reaction fronts (flames and detonations). \hfill

\noindent{\bf Nuclear network \& high density EOS:}
 Nuclear burning is taken into account using Thielemann's  
nuclear reaction-network library (\cite{TNH94},
  and references therein).
The main sources for experimental rates are \cite{CFHZ85},\cite{CF88}
and \cite{WGT90}. Typically, between 20 and 618 isotopes
are taken into account. For the equation of state, we use  a relativistic Fermi-gas  
with  Coulomb corrections and crystallization, and radiation.

\noindent {\bf Opacities and low density equation of state:}
 For low densities, we solve the full rate equations (see below) to determine the level population,
or assume LTE. Typically, about 500 to 600 discrete NLTE levels are included in full NLTE with about 20,000 to
40,000 line transitions. For the 3-D hydro, the effective number of levels is being reduced by about
a factor of 10 by level-merging (\cite{H90}), often referred to as the use of super-levels.
 Complete redistribution over each individual
line both in frequency and in angle are assumed.
 This means that the relative populations
within the sublevels or the merged levels are described by
a Maxwell-Boltzmann distribution.
The data for the atomic line transitions are taken from the new compilation
of Kurucz (\cite{Ku91,Ku95})  and the opacity project (TOPBASE, e.g. \cite{CM92})
from which about 500,000 -2,000,000 lines are extracted,
depending on the temperature and density range.
Photon scattering by free electrons is included in the Klein-Nishina
limit. Free-free cross sections are treated in the hydrogen
approximation with free-free Gaunt factors according to \cite{S60} and \cite{G70}.
  Radiative bound-free cross sections are taken from          
the opacity project (Z$\leq $28).

\noindent{\bf Gamma-ray transport and energy deposition:}
 The $\gamma$-ray transport is computed in  spherical and
 three-dimensions using a  Monte Carlo method (H\"oflich, Khokhlov \& M\"uller 1992, \cite{H02})
including relativistic effects and a consistent treatment of both the
continua and line opacities. 
The interaction processes allowed  are:  
Compton scattering according to the full angle-dependent  Klein-Nishina
formula, pair-production, and bound-free transitions (\cite{AS88}, \cite{BL87}).

\noindent{\bf Radiation transport:} Several modules are available.
 For the spherical geometry and in the stationary case, the radiation transport
equation is solved in the comoving frame (\cite{MKH75}, \cite{MKH76}, \cite{MKH76b}, hereafter MKH-methods).
 For the time dependent cases, we use variable Eddington Tensor solvers (implicit in time) 
(\cite{MM84}, \cite{SM92}, \cite{HMK93}).
 We assume that the Eddington factors are constant during each time step. In the spherical case,
we use the MKH methods. To obtain the correct 
 solution for the Eddington tensor elements in 3-D, we use a Monte Carlo method to compute
the difference between the solution of the non-equilibrium diffusion and full radiation transport equation.
 We  calculate the difference between the solutions for computational accuracy and efficiency.
In particular, the random element in MC avoids spatial fluctuations (and, thus, wiggles) in propagating, plane light fronts
though, still, its speed in the free stream limit may differ from the light speed by 5 to 20 \%.
The same Monte Carlo solver is  used which has been applied  to compute $\gamma$-ray
and  polarization spectra for scattering dominated atmospheres (e.g. \cite{HKM92}, \cite{HHWW01}).
 The Monte Carlo method is appropriate for this problem because of its flexibility 
with respect to the geometrical and velocity structures.

\noindent{\bf AutomaticMeshRefinement for radiation transport by a MC torch:}
 Automatic Mesh Refinement is a well established procedure in hydrodynamics and allows to
adjust the resolution to the requirements. The same holds true also for radiation transport.
 In stellar atmospheres, a logarithmic spacing of the optical depth $\tau$ is adopted to guarantee an appropriate
resolution. In dynamical problems, extended atmospheres/envelopes with arbitrary morphologies
the problem boils down to determine the region of the photosphere, or the 'skin' of an optically
thick object. We start with a grid, equally spaced in mass (in comoving frame) or space (in
 Eulerian frame), and employ a 'Monte Carlo Torch' for the grid refinement. As a basic concept,
  the photosphere is defined as the region from which the photons can escape  or, because of the symmetry of the problem,
 a photon  can penetrate to when coming  the outside.
 With the 'torch',  we eluminate the object from the outside and
calculate the path of the photons (with various energies) in
random directions. If a photon is interacting in a computational cell,
we increase its photon counter. Typically, the number of test photons is of the order of the number of
cells $n_R$ in a grid times the number of representative frequencies $\nu_{rep}$. In our examples (see Figs. \ref{SN99by} 
\& \ref{1dvs3d}), about $10^6$ to $10^7$ are used both in the spherical ($n_r = 912$ \& $\nu_{rep} = 1000$)
) and in the 3-D ($n_r \approx 5  \times 10^5 $ \& $ 10 \times  \nu_{rep}$). Computational
 cells are divided in half if their  photon count exceeds the average by 
a factor of 5 to 10. By eluminating the object from the edges of the domain,
  3-D surfaces   will be traced only if they can be seen from the domain boundaries. Concave structures
such as the inner edges  of Rayleigh Taylor fingers will be missed. Therefore,  $\approx  10$
additional photons are emitted from each grid cell. A  photon triggers
a counter only if it has crossed 2 or more cell boundaries to avoid rezoning
of optically thick layers. The requirements of this approach depends on the physical situation.
 In our examples, we allow for rezoning every 20 to 50 time steps. To test the accuracy, we
increase the number of test photons by a factor of 10 about  every 10 rezoning steps.

\subsection{Accelerated Lambda Iteration \&  Friends}
 The radiation field couples the local, statistical equations
 and the hydro equations, resulting in complex systems of 
 integro-differential equations of rank $DIM = n(x) n(y) n(z) n(\nu) \sum_{el} \sum_{ion(el)}  \sum_{level(el,ion)} n(el,ion,level) $
where n(x,y,z) and  n$(\nu)$ are the spatial and frequency coordinates,  and n(el,ion,level) are the atomic level population of "level" for element "el"
with the ionization "ion". The rank can easily exceed $ 10^{12...13}$ which prohibits a direct solution. However, the iterative solution provides
an effective way to separate global and local quantities.

\noindent a) The method of  accelerated lambda iteration is used
to remove the global dependencies produced by the
the radiation field. 

\noindent b) The statistical equations and the energy terms in the hydro equations
are solved by a partial linearization method
(\cite{MM84}).
 When evaluating the mean intensity, J$_{\nu}$, in the 
rate equations two extreme cases are considered
in which the source function of the specific transition does or
does not dominate the total source function. This approach further assumes
that the variation of the total source function can be represented by the
transition under consideration or by the change of the source 
function during the previous iteration step, respectively. 
The first case is quite similar to the assumptions of  
the equivalent two level approximation, but with the difference that
the perturbation terms are used instead of the total rates.

\noindent c) The concept of ``leading" elements is introduced.
An appropriate ordering of elements or groups of elements
allows  a separation  of the equations, which are solved by 
partial linearization.
  With this technique, the system of 
equations remains small, independently of the problem.  This 
property is needed to achieve numerical stability and computing 
efficiency.

\noindent d) We use the concept of level-locking (\cite{H90}), often called super-levels
 for numerical efficiency. For stability, we use level locking when the system is far from convergence.
 
\noindent e) The scattering and thermal contributions to the source functions
are separated by an equivalent two-level approximation for transitions 
from the ground levels. In effect, this allows to include the non-local nature of optically
thick, scattering dominated transitions already during solution of the full radiation transport problem.
 In effect, this approach 
introduces an acceleration term for the convergence of ALI
(\cite{A72}, \cite{AL92},  \cite{H95}). 

\noindent f) Based on the explicit forms for the total opacity $\chi(\nu)$ and the source function $S(\nu)$,
 we reconstruct redistribution functions in $\nu$ for the photon, 
 and limit the relative change between model iterations  to $\approx 10 \%$.

\noindent {\bf The accelerated lambda iteration:}
The intensity I can be obtained by the solution of the radiation transport equation
\begin{equation}
{\partial I \over {c \partial t}} + Op~I = \chi (S  -  I)
\end{equation}
 where $Op = Op(x,y,z,\delta /\delta x, \delta /\delta z, \delta /\delta z, \delta /\delta \nu$,
and  I is the intensity. On the right hand side, the source function $S$ and and opacity $\chi$ are local quantities.

 Formally, the solution  of the radiation transport or, better, the first momentum $I$                 
can be written as $J(x,y,z,\nu)=\Lambda S(x,y,z,t,\delta/\delta x, \delta/\delta y, \delta/\delta z)$,
and J can be substituted in the rate and energy equation. The matrix elements $\lambda _{j,i}$
describe the contribution of S at the grid point j on J at the grid point i.
 In general, $\lambda $ is not constructed
explicitly. Instead, it can be approximated by the following expression (e.g. \cite{C73}, \cite{S84}, \cite{Hi90},
\cite{H90}, \cite{HL92}, \cite{W91})
\begin{equation}
 J_{\nu}^{(m)}= \Lambda_{\nu}^{(m)} S_{\nu}^{(m)} \approx
\Lambda_{\nu}^{(m-1)} S_{\nu}^{(m-1)} + \Lambda_{\nu}^{* ~(m-1)}
(S_{\nu}^{(m)}-S_{\nu}^{(m-1)})
\end{equation}
 and the solution is obtained iteratively where
$m$ stands for the iteration step. The first term corresponds to the 'classical' $\Lambda$ iteration,
which works well for low optical depths but the convergence of the $\Lambda$ iteration
becomes poor when  $\Lambda $ becomes diagonal and if the scattering part is large. The basic
idea of the ALI is introduce a second, term which removes the local contribution to the radiation 
field and drives the convergence.
 The implementation of the ALI method is reduced to the construction of $\Lambda_*$ and the way how
the expression  for J (eq. 2) is used for the rate  and energy equations.
\begin{figure}[ht]
\vskip -0.2cm
\hskip 2.cm \includegraphics[width=5.2cm,angle=270]{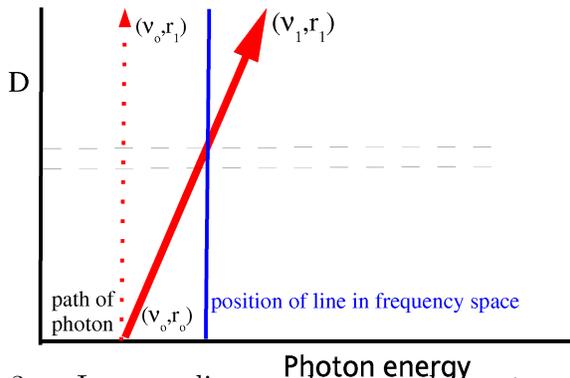}
\vskip -0.2cm
\caption{In expanding envelopes, a photon travels both in the
spatial  and frequency space from $(r_o,\nu _o)$ to $(r_1,\nu _1)$ 
(red arrow).
 Consequently, any line transition (blue line) $\epsilon [\nu_o,\nu_1]$
will effect the absorption probability of the photon whereas, in the static case,
only lines will influence the absorption probability if  $|\nu_{line}-\nu_o|$ is smaller than
the internal line width $\Delta \nu_{line}$. For large velocities,
 a photon can be absorbed by a given line
in  small region with  almost constant physical properties. This allows for an operator splitting of
$Op$ (eq. 1).                   
 }
\label{sob}
\vskip -0.2cm
\end{figure}
\begin{figure}[ht]
\includegraphics[width=9.3cm,angle=90]{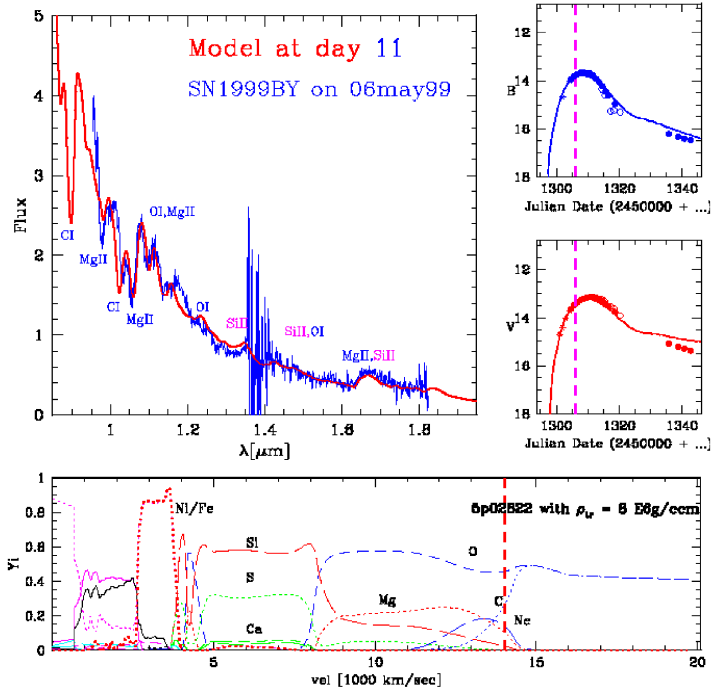}
\vskip -0.54cm
\caption{ Comparison of a explosion model with the SNe~Ia 1999by.
 We show  spectra at day 11 (upper left),
the the B and V light curves  (right plots), and the chemical structure
 (lower panel).
The explosion  and  evolution of the spectra are
calculates self-consistently with the only free parameters being the initial
structure of the exploding White Dwarf, and a parameterized description
of the nuclear burning front.
 The explosion is calculated  by a spherical hydro-modules using 912 depth points
including a  nuclear network with 218 isotopes.
 After about 10 seconds, only radioactive decays are
taken into account but we solve the time-dependent, full NLTE, radiation hydro.
 The LCs
 are based on several thousand  NLTE-spectra utilizing 912 depth, 2000 $\nu $s
 and atoms with  a total of 50 super-levels. For several moments of time,           l,
detailed NLTE spectra have been constructed using 90 depth \& $\approx 30,000 ~ \nu $ points with
$\approx 500$ NLTE levels  (from \cite{HGFS02}).
}
\vskip -0.5cm
\label{SN99by}
\end{figure}
\begin{figure}[ht]
\includegraphics[width=8.0cm,angle=270]{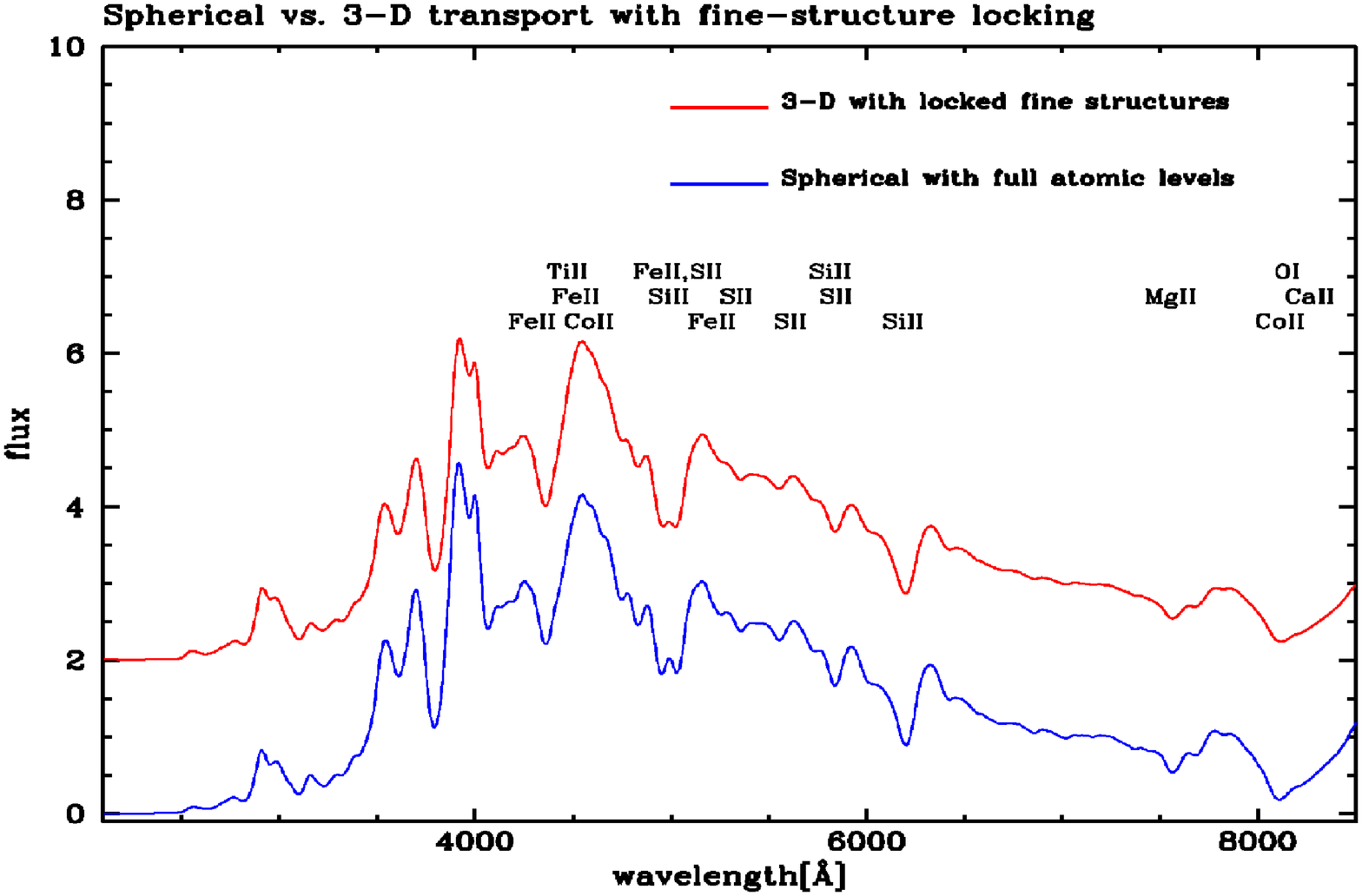}
\vskip -0.04cm
\caption{
 Comparison between theoretical spectra for the SNe~Ia 1999by at about maximum light assuming
spherical symmetry. The spectra are based on our spherical  (blue) and
full 3-D radiation transport scheme, using
90 depth points, 20,000 frequency points and 520 NLTE-levels and 67/67/67 depth points, 2000
frequency points and 50 NLTE-super-levels, respectively. Differences are up to 20 \%. They can be
understood due to the lower resolution in the 3-D calculations. As an application, asphericity effects
of SN99by (see Fig. \ref{SN99by}), see \cite{HHWW01}).
}
\vskip -0.3cm 
\label{1dvs3d}
\end{figure}

\noindent{\bf The diagonal operator for rapidly expanding envelopes:}
 In general, the spatial  and
frequency coordinates are coupled  by the radiation transport equation, and the requirements on the
corresponding resolutions are linked.
 In a Rybicki-like
scheme for solving the radiation transport equation
 (e.g. Mihalas et al. 1975), the frequency spatial resolutions must be better
the internal lines  width, e.g. $\Delta \nu < \nu_{thermal} $, and the velocity difference between
two neighboring cells must be $\Delta v < \Delta \nu /\nu * c$ with c being the speed of light.
 For SN, the velocities are of the order of $10^4 km/sec$, and the line width are given 
thermal velocities $v_{therm}\approx 1km/sec$ which requires a  $\geq 10,000 $ grid points in
each direction. However, the extreme ratio between line width and Doppler shift allows for an
operator splitting of $Op$ (eq. 1) into a directional and frequency dependent part (see Fig. \ref{sob}).
This allows  reduce the solution of the RT problem into the two-stream approximation and
to attribute the frequency shift to the opacity which removes the link between the required spatial and
$\nu $ resolution.

\noindent{\bf Expansion opacities:}
 In the presence of large velocity fields,
the influence of the lines on the opacities
can be well represented in the narrow line limit (\cite{S57},
\cite{C74}) for the given (non-LTE) level populations.
 We use expressions and presumptions (\cite{HMK93})
very similar to those  but but for the comoving frame.
 The Sobolev opacity of a certain line l between a lower and
  upper level i and j, respectively,
 is represented by the expression
\begin{equation}
\chi_l={\pi e^2 \over m_e c } {f_{ij} n_i \over \nu_{ij} \rho} \Biggl(1-
{ g_i n_j \over g_j n_i} \Biggr).
\end{equation}
 
 However, the probability to be absorbed by a given line needs to be modified by the probability
that a photon did interact  before with another line or a continuum. Similar to \cite{K77},
we obtain the following expression if
 N scattering lines are ordered with respect to
 the wavelength
\begin{equation}
\chi_l (\nu) = (\sigma_e+\chi_{\nu}^{bf}+
\chi_{\nu}^{ff})
\eck{1 - \sum^{N_d}_{j
= J} \eck{1 - \exp (-\tau_j)} (\nu_j/\nu)^s \exp \rund{- \sum_{i = 1}^{j
- 1} \tau_i}}^{-1} 
\end{equation}
with
\begin{equation}
\tau_i = \chi_i c \varrho {d r \over d v(r)} ~~and~~
s := (\sigma_e +  \chi_{\nu}^{bf}+\chi_{\nu}^{ff})\cdot c \varrho
{d r \over d
\upsilon}.
\end{equation}

 Note that the Sobolev optical depth is formulated for a given reference frame.
  However, we can transform the opacity from frame 0 to the frame of a neighboring grid point by
\begin{equation}
\chi(v/c)=\chi(0) (1.-2{v \over r \chi_R(0)}) +O(v/c)^2) \approx \chi(0)
\end{equation}
 if we assume
a linear dependence of $\chi_\nu$ in the frequency space and  peace-wise linear velocity fields
(\cite{H90}). Here, $r$ is the local radius of curvature of the velocity field for a direction D.
 For $\Lambda_*$, we can neglect the order O(v/c) in the expression above.

\noindent{\bf The matrix elements of the $\Lambda_*$ operator:}
We construct $\Lambda_*$ in  2-stream approximation, i.e. $Op \rightarrow \delta /\delta D$ and
use intensity like variables $u = I^++I^-$ for direction D with the grid d. As       in static atmospheres (\cite{OAB86}), we
obtain
\begin{equation}
{\delta I \over \delta d} = \chi (I - S) ~~and~with~\delta \tau = \chi \delta d  \Rightarrow {\delta \over \delta \tau} u = u -S
\end{equation}
By using the flux and intensity like variables we  get the final
equation
\begin{equation}
{{u_{d + 1}\over \Delta \tau_d \cdot \Delta \tau_{d + 1/2}} - u
_d \rund{{1 \over \Delta \tau_{d} \cdot \Delta \tau_{d + 1/2}}
+ {1 \over \Delta \tau_d \cdot \tau_{d - 1/2}}} + {u _{d - 1}\over
\Delta \tau_{d - 1/2} \Delta \tau_d}}   
 = u_d - S_d.
\end{equation}
 with 
\begin{equation}
 S_d = 1,
u^*_{d - 1} = u_d^* \exp \eck{- (\tau_d - \tau_{d - 1})} ,
u^*_{d + 1} = u_d^* \exp \eck{- (\tau_{d + 1} - \tau_d)} 
\end{equation}
\begin{equation}
 \chi_{d \pm 1/2} : = {1 \over 2} \eck{\chi_{d \pm 1} + \chi_d}
 \Delta \tau_{1 + 1/2} : = \chi_{d + 1/2}  \eck{z_d - z_{d + 1}},
\end{equation}
where $u^*_{d}$ is the "flux" like quantity. After some transformations
one gets the algebraic expression
\begin{equation}
\Delta \tau_{d - 1/2} : = \chi_{d - 1/2} \eck{z_{d - 1} - z_d},
\Delta \tau_{d + 1/2}
: = {1 \over 2} \eck{\Delta \tau_{d + 1/2} + \Delta \tau_{d -
1/2}} 
\end{equation}
\begin{equation}
u^*_d =  \eck{1 + \rund{{1 - \exp ( - \Delta \tau_{d}) \over
\Delta \tau_d \Delta \tau_{d + 1/2}} + {1 - \exp ( - \Delta \tau_{d -
1})
\over \Delta \tau_d \cdot \Delta \tau_{d - 1/2}}}} ^{-1}
\end{equation}
 At the inner boundary of each ray we take for the optically thick and thin cores
$u^*_{d = d_{max}} = 1$
$u^*_{d =d_{max}} = (1 - e^{- \Delta \tau_{d=d_{max}}})  $, respectively.
From the the known u$_d^*$ we get the diagonal element by
\begin{equation}
\Lambda_{\nu}^* = \int\limits^*_0 d \mu d\phi u^*_d. 
\end{equation}

 One remaining problem related to the property that the locality of the operator
depends on the resolution, i.e. the convergence rate becomes grid dependent and,
for high resolutions, the system may become unstable. The information on corrections  propagates
 only one grid point per model interaction m. To overcome these problems and if the solution is far from convergence,
   we refine the
the 'grid/average over several zones' corresponding to the 
thermalization depth $\tau_{therm}$ of the transition with the highest opacity of a given element. The estimate of $\tau_{therm}$
 based on the equivalent-two-level approach. Successively, we  increase the
effective resolution to the full grid resolution. For our applications,  this approach has been found to be  more stable
rather than using off-diagonal elements (\cite{W91}) but, in light of discussions at this meeting, the latter approach will be revisited.
 
\noindent{\bf The matrix elements of the statistical equations:}
\noindent
 The matrix elements of the statistical equation A x = b of each species are
given by
 
\begin{equation}
A_{ij} = \cases{- (R_{ij} + C_{ij}) &$i < j$ \cr
- ({n_i^*\over n^*_j}) (R_{ji} + C_{ij}) &$j > i$ \cr
\sum^{i - 1}_{l=1} ({n_l^*\over n_i^*})
 ( R_{il} + C_{il}) + \sum^{n}_{l = i + 1}
(R_{il} + C_{il}) &$j = i$ \cr} 
\end{equation}
$R_{ij} $ and  $C_{ij}$
 are the radiative and collisional rates relating the levels i and j.
 
The radiative rates between a lower and upper level i and j,
respectively, are given by
\begin{equation}
  R_{ij} = 4 \pi \int
{\alpha_{ij} (\nu) \over h \nu} J_\nu d_\nu + R(\gamma-rays)
\end{equation}
\begin{equation}
R_{ji} = 4 \pi  \int
{\alpha_{ij}(\nu) \over h \nu }
\eck{{2 h \nu^3 \over c^2} + J_\nu} e^{-{h \nu \over k T}} d\nu .
\end{equation}
 
 Here, where the coupling between the radiation field and the
statistical equations occurs, in principle, the solution of
the statistical equations of the current model iteration (m) has
 to be known in order  to calculate the
new value of the radiation field
$J_\nu^{(m)}$.
In order to estimate the updated rates, we distinguish  two cases:

\noindent{Case A:}
$S_\nu/\chi_\nu \gg S_{ij}/\chi_{ij}$ and B)
$S_\nu/\chi_{\nu}
  \approx S_{ij}/\chi_{ij}
 $.
\noindent{Case A:} Eqs. 15 and 16 can be
         written   as      a sum  of the "zero" order
\begin{equation}
R_{ij} = 4 \pi \int {\alpha_{ i j}(\nu) \over h \nu} \Lambda_\nu
S_\nu^{(m - 1)} d \nu
\end{equation}
\begin{equation}
R_{ji} = 4 \pi \int {\alpha_{ i j}(\nu) \over h \nu}
\Biggl\lbrack
{2 h \nu^3 \over c^2}+ \Lambda_\nu
 S^{(m-1)
}_\nu
\Biggr \rbrack e^{{-h \nu \over k T}} d~ \nu
 d \nu
\end{equation}
 
\noindent
and by correction terms of  "first"  order
\begin{equation}
R_{ij}^{corr} =
 4 \pi \int {\alpha_{i j} (\nu)\over h \nu} \Lambda_\nu^*
  (S^{(m)}_\nu - S^{(m - 1)}_\nu) d\nu 
\end{equation}
\begin{equation}
R_{ji}^{corr} =
 4 \pi \int {\alpha_{i j}(\nu) \over h \nu}
 \Lambda_\nu^*  (S^{(m)}_\nu - S^{(m - 1)}_\nu) e^{{-h \nu \over k T}} d~ \nu
\end{equation}
 
 In order to  estimate  the source functions
 we  use again the concept of the
"leading" elements.
 If the total emissivity at the considered frequency
is dominated by an element which is higher in the hierarchy, the new
difference between the old and the new source function is
substituted by the difference of the dominating transitions
(marked by the index {\sl dom}), i.e.
 
\begin{equation}
 (S_\nu^{(m)} -S_\nu^{(m-1)}) \approx
  (S_{dom}^{(m)} -S_{dom}^{(m-1)}) \cdot \chi_{dom}/\chi_{\nu}.
\end{equation}
 
 Otherwise the monochromatic source function is
just an extrapolation in the second
order of the iteration steps, i.e.
 
\begin{equation}
\Delta S_\nu^{(m)} \simeq \Delta S_\nu^{(m - 1)} + {\Delta S^{(m
- 2)} -
\Delta S^{(m - 1)} \over  2} + ... 
\end{equation}
with
\begin{equation}
\Delta S_{\nu}^{(m)} := S_{\nu}^{(m)} -S_{\nu}^{(m-1)}.
\end{equation}
 
~
 
\noindent
{\sl Case B:} If the  transition in consideration dominates the total
source function, we use the approach
$ S_{\nu} \approx S_{ij} (\nu).$
 In order to get expressions for the "first" order corrections to the
rates, essentially,  using
the same assumptions as in the case of the
 equivalent two level approach.
The explicit forms of the source functions 
allow us to substitute the departure coefficient of the
lower level in the rate equations (e.g. Pauldrach 1987). After some simple transformations,
 one ends up with the expression
for the first order correction of the rates
\begin{equation}
R_{ij}^{corr} = 0 ~~~and~~~
R_{ji}^{corr} = 4 \pi \int {\alpha_{i j} (\nu) \over h
\nu} \Lambda_{ij}^*(\nu) {(S_{ij}^{(m-1)}(\nu)
 - S^{(m)}_{ij} (\nu))\over S_{
ij }^{(m)}(\nu)} d \nu  
\end{equation}
 
The separation between  the cases A and B is needed to
overcome numerical instabilities during the model iteration.  
 The rate equations and the hydrodynamics are iteratively coupled via the energy equation (e.g. \cite{HMK93}).
For $J$, we use the same expressions and case separation as just outlined.

\subsection{The Time-dependent Rate Equations}
An approximate solution of
the time dependent non-LTE
problem is outlined (\cite{H91b}).
 The statistical equations are reduced to
a form of time independent
equations which can be solved by one of the standard methods
for the  time independent case, where the time dependent
quantities enter the statistical equations just in an additional
explicit rate.
 These
are calculated analytically solving
a linear inhomogeneous
differential equation of first order.
 
We assume  that the time dependence is caused by a limited number of transitions.
 In our implementation, we assume that the time dependence is caused by the bound-free transitions.
This condition  is valid as long as 
forbidden transitions are  suppressed because allowed bound-bound
transitions are faster than  bound-free transitions by orders of
magnitude. This limits the applicability e.g., for SNeIa,   to  times $\leq$ 40 to 60 days after
the explosion. Forbidden lines causes the discrepancies in the late LCs in Fig. \ref{SN99by}.
 
In general, the time dependent rate equations are given by the expression
\begin{equation}
{\partial n_i \over \partial t} + \nabla (n_{i} \underline{v}) =
\sum_{i\ne j}
 (n_{j} P_{ji} - n_{i} P_{ij}) + n_{k} (R_{ki} + C_{ki}) - n_{i} (R_{ik} +
C_{ik})
\end{equation}
where the rate P$_{ij}$ is the sum of radiative and collisional processes.

The time
independent solution $\tilde n_{i}$ of the statistical equations is
defined by the equations
\begin{equation}
  0 = \sum_{i
\ne j} (\tilde n_j P_{ji} - \tilde n_i P_{ij}) + \tilde n_k (R_{ki} + C_{ki}) - \tilde n_i (R_{ik} +
C_{ik}) 
\end{equation}
that can be solved by a standard method.
 
 For most astrophysical
applications, the time scales of allowed bound-bound
transitions are much shorter than those of
 the bound-free transitions, and 
we obtain the relation
\begin{equation}
\Rightarrow {n_i \over n_j} = {\tilde n_i \over \tilde n_j}.
\end{equation}
 
Subtracting the time independent rate equations from
 the time dependent ones both of which
are normalized to $n_i$ and $\tilde n_i$, respectively, results in
\begin{equation}
{\partial n_i \over \partial t} + {\partial \over \partial r} (n_i
v(r)) =
 n_k  (R_{ki} + C_{ki}) - {\tilde{n_k} \over
\tilde{n_i}} (R_{ki} + C_{ki})~n_i
\end{equation}
i.e. an inhomogeneous linear differential equation of  first order
\begin{equation}
\Rightarrow {d n_i \over d t} = - C_1 n_i + C_2 
\end{equation}
with
\begin{equation}
 C_1 : = {\tilde{n_k} \over \tilde{n_i}} (R_{ki} + C_{ki}) ~~and~~
 C_2 : = n_k (R_{ki} + C_{ki}) . 
\end{equation}
 This inhomogeneous ODE is   solved   analytically as an initial boundary problem. If more than
two ionization stage are important, we end up with a system of ODEs to be solved  numerically.
 $n_k$ and $\tilde{n_k}$ are determined iteratively by using
charge conservation in a simple fix point iteration, or a Newton-Raphson method.
\begin{acknowledgments}
This research is  supported in part by  NASA Grant LSTA-98-022.
\end{acknowledgments}

\end{document}